\documentclass[a4paper]{article}

\usepackage[]{geometry} 
\usepackage{tikz}
\usetikzlibrary{arrows.meta, positioning, quotes}
\usepackage{amsmath}
\usepackage{amsthm}
\usepackage{amssymb}
\usepackage{float}
\usepackage{enumerate}
\usepackage[utf8]{inputenc}
\usepackage{hyperref}
\hypersetup{
	unicode,
	pdftitle={Granger Test for heteroskedastic and anomalous data using generalized least squares},
	pdfproducer={LaTeX},
	pdfcreator={pdflatex}
}

\usepackage{comment}
\usepackage{enumitem}

\theoremstyle{plain}
\newtheorem{theorem}{Theorem}

\theoremstyle{definition}
\newtheorem{definition}[theorem]{Definition}
\newtheorem{example}[theorem]{Example}

\newtheorem{problem}[theorem]{Problem}

\newtheorem{property}[theorem]{Property}
\newtheorem{observation}[theorem]{Observation}
\newtheorem{method}[theorem]{Method}

\newcommand{\x}{\textbf{x}}
\newcommand{\y}{\textbf{y}}
\newcommand{\cov}{\text{cov}}

\usepackage{graphicx, color}

\usepackage{algorithm, algpseudocode} 
\usepackage{mathrsfs} 
\algnewcommand\algorithmicforeach{\textbf{for each}}
\algdef{S}[FOR]{ForEach}[1]{\algorithmicforeach\ #1\ \algorithmicdo}

\title{Granger causality test for heteroskedastic and structural-break time series using generalized least squares}
\author{Hugo J. Bello \footnote{\texttt{hugojose.bello@uva.es} Department of Applied Mathematics, University of Valladolid (Campus Soria)}}

\begin{document}
\maketitle

\begin{abstract}
	This paper proposes a novel method (GLS Granger test) to determine causal relationships between time series based on the 
	estimation of the autocovariance matrix 
	and generalized least squares. We show the effectiveness of proposed autocovariance matrix estimator 
	(the sliding autocovariance matrix) and we compare the proposed 
	method with the classical Granger F-test with via a synthetic dataset and a real dataset composed by cryptocurrencies. The simulations 
	show that the proposed GLS Granger test captures causality more accurately than Granger F-tests 
	in the cases of heteroskedastic or structural-break residuals.
	Finally, we use the proposed method to unravel unknown causal relationships between cryptocurrencies.

\end{abstract}

\section{Introduction}
\label{sec:intro}

Granger causality is a statistical concept that helps us determine whether the information in one time series is useful in predicting another.
It is widely used in numerous fields, including economics, finance, geology, genetics and neuroscience, 
to understand the relationships between variables and to identify possible causal connections.
 
The main tool for studying time series causality is the Granger causality F-test (\cite{granger1969investigating}, \cite{granger1980testing}) 
which is based on ordinary least squares (OLS) 
estimation and therefore is tied to the assumptions of OLS (among them homocedasticity of the residuals)

In many real-world applications, we are confronted
with time series suffering from a variety of problems such as heteroscedasticity (non-equal variance amongst time observations)
 or structural breaks
(the appearance of two or more stationary periods with different means). 
In these situations OLS is not recommended and Granger F-test can 
be inadequate. 

The aim of this paper is to present a method for determining Granger causality based on generalized least squares based on the 
estimation of the autocovariance matrix of the residuals.

\subsection{Preliminaries}
In this section we will present introducing notation and definitions. 
We also review the notion of \emph{Granger causality}.

\subsubsection{Granger causality F-test}\label{sec_granger_ftest}

The Granger causality (\cite{granger1969investigating}, \cite{granger1980testing}) test is a statistical hypothesis test for 
determining whether one time series $\x = (x_t)^N_{t=0}$ is useful in forecasting another $ y = (y_t)^N_{t=0}$. 
In particular Granger causality focuses on the possibility of $\x$ and $\y$ to predict future values 
of $\y$ \emph{significatively better} than those of $\y$ alone. In this case it is said that $\x$ Granger causes $\y$

The idea behind this is that a cause should be helpful in predicting the future effects, beyond what can be 
predicted solely based on their own past values.

The test null hypothesis states that the linear regression model

\begin{equation}\label{eq_granger_rm}
y_t = \beta_0 + \sum^{p}_{k=1} \beta_k y_{t-k} + \varepsilon_t
\end{equation}

approximates $\y$ significatively better than the model

\begin{equation}\label{eq_granger_ur}
	y_t = \beta_0 + \sum^{p}_{k=1} \beta_k y_{t-k} + \sum^{p}_{k=1} \beta_k' x_{t-k} + \varepsilon_t
\end{equation}

If the null hypothesis is correct, this will imply that the lagged values of $\x$ add explanatory power to the prediction of $\y$ and therefore 
the process behind $\x$ causes $\y$.

To test if model (\ref{eq_granger_ur}) is significatively more accurate than (\ref{eq_granger_rm}) the following statistic is used 
 
\begin{equation}\label{eq_granger_f}
	\frac{(SSR_{RM} - SSR_{UM})/N}{SSE_{UM}/(N-2)(p-1)}
\end{equation}

where   $SSR_{RM}$ and $SSR_{UM}$ are the sum of residuals for the restricted and unrestricted models respectively

\[SSR_{RM} = \Sigma_t (y_t - \beta_0 - \sum^{p}_{k=1} \beta_k y_{t-k})^2
\]
\[
SSR_{UM} = \Sigma_t (y_t - \beta_0 - \sum^{p}_{k=1} \beta_k y_{t-k} - \sum^{p}_{k=1} \beta_k' x_{t-k})^2
\]

If the necessary assumptions for ordinary least of squares are satisfied,  (\ref{eq_granger_f}) follows a $F(p, N-2p-1)$ distribution under the 
null hypothesis.

\begin{observation} \label{obs_backward}
	Notice that \ref{eq_granger_rm} can be understood as writing 
	$\y$ as a linear combination of the lagged time series $B^k \y$ (using the 
	backward operatior notation), that is:
	\[\y = \beta_0 + \sum^{L}_{k=1} \beta_k B^k  \y + \varepsilon\]
	
	Similarly \ref{eq_granger_rm}
	\[\y = \beta_0 + \sum^{L}_{k=1} \beta_k B^k \y + \sum^{L}_{k=1} \beta_k  B^k \x + \varepsilon\]

	Therefore Granger causality can be understood as the use of 
	a F-test to compare two multilinear regression models.
\end{observation}

\subsubsection{Generalized Least Squares}

For a linear regression model of the form 

\begin{equation}\label{eq_lin_mod}
	{\displaystyle y_{i}=\beta _{1}\ x_{i1}+\beta _{2}\ x_{i2}+\cdots +\beta _{p}\ x_{ip}+\varepsilon _{i},}
\end{equation}

The following assumptions must be satisfied for ordinary least squares (OLS) to have the desired asymptotic properties:

\begin{itemize}[noitemsep]
\item[(P1)] \textbf{Correct specification}. Te underlying process generating the data must be in esence linear.
\item[(P2)] \textbf{Strict exogeneity}. The errors in the regression must have conditional mean zero: ${\displaystyle \operatorname {E} [\,\varepsilon \mid X\,]=0.}$.
\item[(P3)] \textbf{No linear dependence}. The regressors must all be linearly independent.
\item[(P4)] \textbf{Homoscedasticity}  $E[\varepsilon_i^2 | X]= \sigma^2 \;\forall i $. The error term has the same variance $\sigma^2$ in each observation.
\item[(P5)] \textbf{No autocorrelation} $E[\varepsilon_i \varepsilon_j | X]= 0 \;\forall i\not= j$. The errors are uncorrelated between observations.
\item[(P6)] \textbf{Normality.} It is sometimes additionally assumed that the errors have normal distribution conditional on the regressors
\end{itemize}

If homocedasticity (P4)  or non-autocorrelation (P5) assumptions are not satisfied we can use \textbf{Generalized Least Squares} (GLS) to better approximate 
the parameters of (\ref{eq_lin_mod}). 

Using the standard matricial notation (\ref{eq_lin_mod}) can be written as 

\[
\y = X \boldsymbol \beta + \varepsilon
\]

where $\y = (y_i)$, $X = (x_1^T \ldots x_p^T)$ is the design matrix and $\varepsilon$ is the error term.
if the error term satisfies 
that $E[\varepsilon | X] = 0 $ and denoting  $\cov [\varepsilon | X] = \Omega$ the non-singular covariance matrix 
of the residuals.
the GLS estimate for $\boldsymbol \beta$ is 

\begin{equation}\label{eq_gls}
\widehat{\boldsymbol \beta}_{GLS} = \text{argmin}_{\boldsymbol \beta} (y - X \boldsymbol \beta)^T \Omega (y - X \boldsymbol \beta)
=\left(\mathbf {X} ^{\mathsf {T}} \Omega^{-1}\mathbf {X} \right)^{-1}\mathbf {X} ^{\mathsf {T}} \Omega   ^{-1}\mathbf {y} 
\end{equation}

(see \cite[\S 9.3]{greene2003econometric}) Notice that for $\Omega = \sigma^2 I$ (where  $I$ is the identity matrix) 
we are under the assumptions of OLS and the resulting estimator is

\begin{equation}\label{eq_ols}
	\widehat {\boldsymbol \beta}_{OLS} = \left(\mathbf {X} ^{\mathsf {T}}\mathbf {X} \right)^{-1}\mathbf {X} ^{\mathsf {T}}\mathbf {y} 
\end{equation}

It is known that 
\begin{align}\label{eq_var_gls}
	E[\widehat{\boldsymbol \beta}_{GLS}] &= \boldsymbol \beta_{GLS}\\
	\cov[\widehat{\boldsymbol \beta}_{GLS}] &= V = (X^T \Omega^{-1} X)^{-1} \label{eq_var_gls}
\end{align}
In fact 
\[\sqrt{N} (\widehat{\boldsymbol \beta}_{GLS} - \boldsymbol \beta_{GLS}) \longrightarrow^D N(0, V)\]

\subsubsection{Wald Test}
The Wald test is a statistical hypothesis test that assesses constrains on statistical parameters for regression models based on the weighted distance 
between an unrestricted estimate and its hypothesized value under the null hypothesis (see \cite[\S 5.3]{greene2003econometric}).

Let $\widehat{\boldsymbol \beta}$ the sample estimate for the regression model (GLS or OLS) model with covariance $V$ as  described before.
If $Q$ hypothesis on the $p$ parameters are expressed in the form of a $Q\times p$ matrix $R$:

\begin{align*}
&H_0: R\beta = r \\
&H_1: R\beta \not = r
\end{align*}

The wald test statistic is 
\begin{equation}\label{eq_wald}
\displaystyle (R\widehat{\boldsymbol \beta} - r)^T \cdot \bigg(R\widehat V  R^T \cdot \frac{1}{n}\bigg)^{-1} \cdot (R\widehat{\boldsymbol \beta} - r)
\end{equation}

Under the null hypothesis, the Wald statistic (\ref{eq_wald}) follows a $F(Q, N-p)$ distribution.

\subsubsection*{Granger F-test as a Wald Test}
The Granger F-Test described in (\ref{eq_granger_f}) is in fact a particular case of the Wald test. 
If we consider $\widehat{\boldsymbol \beta}_{OLS}$, then we can stablish the unrestricted regression model 

\begin{equation}\label{eq_granger_wald}
\y = \beta_0 + \beta_1 B \y + \ldots + \beta_p B^p \y + \beta_1' B \x + \ldots + \beta_p' B^p \x
\end{equation}
Where $B$ is the backshift operator $B^k x = (x_{t-k})_{t}$.
So for $y$ to be caused by $x$ the coefficients $\beta_1', \ldots \beta_p'$ must be zero, therefore we need to test the 
hypothesis 

\begin{align}
	H_0: &\beta_k' = 0 \text{ for all } k\leq p\label{granger_hip_test}\\
	H_1: &\beta_k' \not = 0 \text{ for some } k\leq p\nonumber
\end{align}

Defining the matrices

\[R = \begin{pmatrix}
0 &       &   &   &       &  &  \\
  & \ddots&   &   &       &  &  \\
  &       & 0 &   &       &  &  \\
  &       &   & 1 &       &  &  \\
  &       &   &   & \ddots&  &  \\
  &       &   &   &       & 1&  \\
\end{pmatrix};
	\; \; \boldsymbol \beta = \begin{pmatrix}
		\beta_1   \\
		\vdots    \\
		\beta_p   \\
		\beta_1'  \\
		\vdots    \\
		\beta_p'       
		\end{pmatrix}; \; \; r = \begin{pmatrix}
			0        \\
			  \vdots \\
			0        
			\end{pmatrix}
\]
We can codify the test \ref{granger_hip_test} as 

\begin{align}
	H_0: &R\boldsymbol \beta  = 0  \label{granger_hip_test2}\\
	H_1: &R\boldsymbol \beta \not = 0 \nonumber
\end{align}

It can be shown \cite[\S 5.4]{greene2003econometric}, that with this notation the corresponding wald statistic (\ref{eq_wald}) 
in fact coincides with the F-test statistic (\ref{eq_granger_f})

\section{Methodology}

The Granger F-test (\ref{eq_granger_f}) assumes that 
conditions (\ref{eq_lin_mod}) are satisfied. We aim  
to present a version of the Granger test based on generalized least 
squares, for that we need $\widehat \Omega$, an estimate for the 
 covariance matrix $\cov [\varepsilon | X] = \Omega$. 
 
In most cases $\Omega$ is not known and a reasonable approach is to use the $\beta_{OLS}$ to obtain
the residuals 

\[
r_t(\beta) = y_t - \beta_0 - \sum^{p}_{k=1} y_{t-k} \beta_k  - \sum^{p}_{k=1} x_{t-k} \beta_k' 
\]

And attempt to estimate $\Omega$ using the covariance matrix of $r_t$. This procedure is often called \textbf{feasible least squares}.

The difficulty lies  in the fact that the covariance matrix of a time series (which is often called \emph{autocovariance matrix})
is not known in general. To overcome this problem in many 
cases $r_t$ is assumed to follow a known model such as AR(1), whose theoretical autocovariance matrix is known and can be
obtained  from the model parameters. This approach is very restrictive since in general $r_t$ can take many forms, for this reason we will first 
tackle the following problem:

\begin{problem}
Given a time series $\x = (x_t)$ how 
can we estimate the covariance matrix $\Omega = (\cov(x_t, x_{t'}))_{t, t'}$?
\end{problem}

Since in general the previous problem can be really difficult to tackle we will impose 
certain assumptions. We will focus on the following realm of very general time series: \textbf{the locally jointly-stationary}
which as we see admit a convenient estimation of their covariance (which we will call the \textbf{the sliding autocovariance matrix})

\subsection{Locally jointly-stationarity and the sliding autocovariance matrix}

\begin{definition}\label{def_cross}
A time series $\x = (x_t)$ is \textbf{locally jointly-stationary} if there exists an increasing sequence of time instances
$0 < t_1 < t_2 < \ldots < t_n$ (called \textbf{time breaks}) such that each subsequence
\[x^{(k)} = x_{t_{k}: t_{k+1}} = \{ x_t : t\in [t_k, t_{k+1}]\}\] 
is stationary and jointly-stationary with respect to the rest of the subsequences.

Recall that two time series $\x, \y$ are jointly-stationary if they satisfy $\cov(x_t, y_t) = \cov(x_{t+h}, y_{t+h})$
\end{definition}

\begin{example}
Stationary time series are locally jointly-stationary. 
This is very easy to verify since every subseries of a stationary time series 
will be cross stationary with any other subseries. One can consider any instance $T$ and the initial subseries $x_{0: T}$ 
is trivially cross stationary with the rest of the series by definition.
\end{example}

\begin{example} A time series $(x_t)$ is called \textbf{stationary with structural breaks} if it satisfies 
	\[
	x_t = \alpha + \delta D_t + \varepsilon_t
	\]
	where 

	\[
	D_t = \begin{cases}
		1 & \text{ if } t\geq T_B + 1\\
		0 & \text{ otherwise }\\
	\end{cases}
	\]
	for $\alpha, \delta\in \mathbb R$ and $\varepsilon_t$ stationary. 
These time series were introduced by Perron (see \cite{perron1989great} and \cite{lee2001testing}).
\end{example}

\begin{property}\label{prop_structural}
Stationary time series with structural breaks are locally jointly-stationary. 
\end{property}

\begin{proof}
Consider the subsampled time series 
$a_t =  \alpha + \varepsilon_t$ defined for values of $t$ between $0$ and $T_b$, and $b_t= \alpha + \delta + \varepsilon_t$. 
Since $\varepsilon_t$ is stationary, this two time series are jointly-stationary:
\begin{align*}
\cov(a_{t_1}, b_{t_2}) &= \cov(\alpha + \varepsilon_{t_1}, \alpha + \delta + \varepsilon_{t_1})\\
&= \cov(\varepsilon_{t_1}, \varepsilon_{t_2}) = \cov(\varepsilon_{t_1 + h}, \varepsilon_{t_2 + h})\\
&= \cov(\alpha + \varepsilon_{t_1 + h}, \alpha + \delta + \varepsilon_{t_1 + h})\\
&=\cov(a_{t_1+h}, b_{t_2+h}) 
\end{align*}

Therefore $\x$ satisfies definition \ref{def_cross}, taking the partition $0\leq t\leq T_b$ and $T_b<t$.
\end{proof}

\begin{definition}
Given a time series $\x = (x_t)$, we define \textbf{the window of length} $\tau$ at $t_0$ as
\[w_\tau(\x, t_0) = x_{t_0: t_0 - \tau} = \{ x_t : t\in [t_0-\tau, t_0]\}\]
\end{definition}

\begin{definition}
Let $\x = (x_t)$ be a time series, and two time instants $t_1, t_2$. Consider the windows of length $\tau$

\begin{align*}
w^1 &= w_\tau (\x, t_1)\\
w^2 &= w_\tau (\x, t_2)	
\end{align*}

We will call the \textbf{windowed sample autocovariance} of length $\tau$ at $t_1, t_2$ to 

\begin{align}
	\widehat \gamma_{\tau} (t, t')  &= \displaystyle{\widehat \gamma_{w^1 w^2} (t, t')}\\
	&= \frac{1}{\tau}  \sum^\tau_{k=0} (w^1_{k} - \overline {w^1})(w^2_{k} - \overline {w^2})\label{def_windowed_sample}\\
	&= \frac{1}{\tau}  \sum^\tau_{k=0} (x_{t-k} - \overline {w^1})(x_{t'-k} - \overline {w^2})
\end{align}
 
which coincides with the sample cross-covariance
\footnote{The sample cross-covariance of two (jointly-stationary) time series $\x$ and $\y$ is defined as 
	$\widehat\gamma_{xy}(h) =\sum (x_{t+h} -\overline{x})(y_{t+h} -\overline{y})$. See example 1.23 \cite{shumway2000time}}
between the time series $w_\tau(\x, t)$ and  $w_\tau(\x, t')$ 

\end{definition}

\begin{property}\label{proper_estimator}
Let  $\x = (x_t)$ be a locally jointly-stationary time series  with time breaks $0 < t_1 < t_2 < \ldots < t_n$.
Given $t, t'$, taking 
\[\tau = \displaystyle{\underset{\underset{ T=t,t'}{1\leq m \leq n}}{\operatorname{argmin}}} |T-t_m|\] 
the windowed sample autocovariance of length $\tau$ is an estimator for $\cov(x_{t}, x_{t'})$
\end{property}

\begin{proof}
Suppose that the time break immediately lower that $t$ is $t_m$ and that the one immediately lower than $t'$ is $t_{m'}$. 
We will consider first the case that $t_m$ and $t_{m}'$ are different. 

Notice that (following the notation in \ref{def_cross}), by the choice of $\tau$

\begin{align*}
w_\tau (\x, t, \tau)  &= 	w_\tau (\x^{(t_m)}, t, \tau)\\
w_\tau (\x, t', \tau) &= 	w_\tau (\x^{(t_{m'})}, t', \tau)
\end{align*}

Therefore, $\widehat \gamma_\tau (t, t')$, the windowed sample autocovariance of length $\tau$ coincides with the sample cross covariance of the previous two subseries windows
$w_\tau (\x^{(t_m)}, t, \tau), w_\tau (\x^{(t_{m'})}, t', \tau)$.

Since the subsequences  $\x^{(t_m)}$ and $\x^{(t_{m'})}$ are jointly-stationary, $\widehat \gamma_\tau (t, t')$ 
estimates the covariance 
\[\cov(w_\tau (\x^{(t_m)}, t, \tau), w_\tau (\x^{(t_{m'})}, t', \tau))\] 
which must coincide with 
$\cov(\x^{(t_m)}_t, \x^{(t_{m'})}_{t'}) = \cov(x_{t}, x_{t'})$. 

If $t_m = t_{m}'$ then in the previous argument  $\x^{(t_m)} = \x^{(t_{m'})}$ 
and the same consequence follows using the stationarity of $\x^{(t_m)}$.

\end{proof}

\begin{observation}
Notice that since the sample cross-covariance for jointly-stationary time series is a biased estimator, 
the windowed sample autocovariance $\widehat \gamma_{\tau} (t, t')$ is a biased estimator. Nevertheless, in the case that the 
expected value $E[\x]$ is known, replacing the average by the expected value in the formula the cross-covariance becomes unbiased 
and therefore the same holds for $\widehat \gamma_{\tau} (t, t')$ in view of the previous proof.
\end{observation}

\begin{definition}
The \textbf{sliding autocovariance matrix} of length $\tau$ is the $N\times N$ matrix  $\Omega_\tau$ defined
thewindowed sample autocovariance $\widehat \gamma_{\tau} (t, t')$ for every pair of time instances, that is 

\begin{equation}\label{eq_sliding_autocov_matrix}
\Omega_\tau = \displaystyle{(\widehat \gamma_{\tau} (t, t'))_{t, t'}}
\end{equation}

By prop. \ref{proper_estimator}  $\Omega_\tau$ is an estimator for the covariance matrix $\Omega$ in the case that the series $\x$ is 
locally jointly-stationary. By prop. \ref{prop_structural} if $\x$ is stationary with structural breaks,  $\Omega_\tau$ estimates $\Omega$.
\end{definition}

\begin{observation} \label{ref_obs_complete}
	Notice that in (\ref{eq_sliding_autocov_matrix}) for low values of $t, t'$ the estimation $\widehat \gamma_{\tau} (t, t'))_{t, t'}$ becomes 
	imprecise due to the small number of values before. One way to fix this problem in certain situations is to complete the time series with 
	values before $0$ using $x_{-t} = x_t$.
\end{observation}

\begin{example}
	Consider time series $x_t = \phi_1  x_{t-1} + \varepsilon_t$ with $\phi_1 = 0.9$. 
	Since the time series follows an AR(1) model, we know that 
	the autocovariance 
	
	\begin{equation}\label{eq_cov_ar1}
		\cov(x_t, x_{t+h}) = \phi_1^h \cdot var(x_t)
	\end{equation}
 	
	Figure \ref{fig_windowed} shows a simulation of this time seres at the top, 
	the theoretical covariance matrix using the previous formula \ref{eq_cov_ar1} is shown on the left side and the sliding covariance matrix 
	estimated using (\ref{eq_sliding_autocov_matrix}) is shown on the left. For the imprecision around low values we used the procedure described 
	in obs. \ref{ref_obs_complete}. To calculate the autocovariance matrix we used the value $\tau = N/3$ where $N$ is the sample size for the time 
	series, lower values of $\tau$ produce similar estimations.

	\begin{figure}[H]
		\centering
		\includegraphics[scale=0.35]{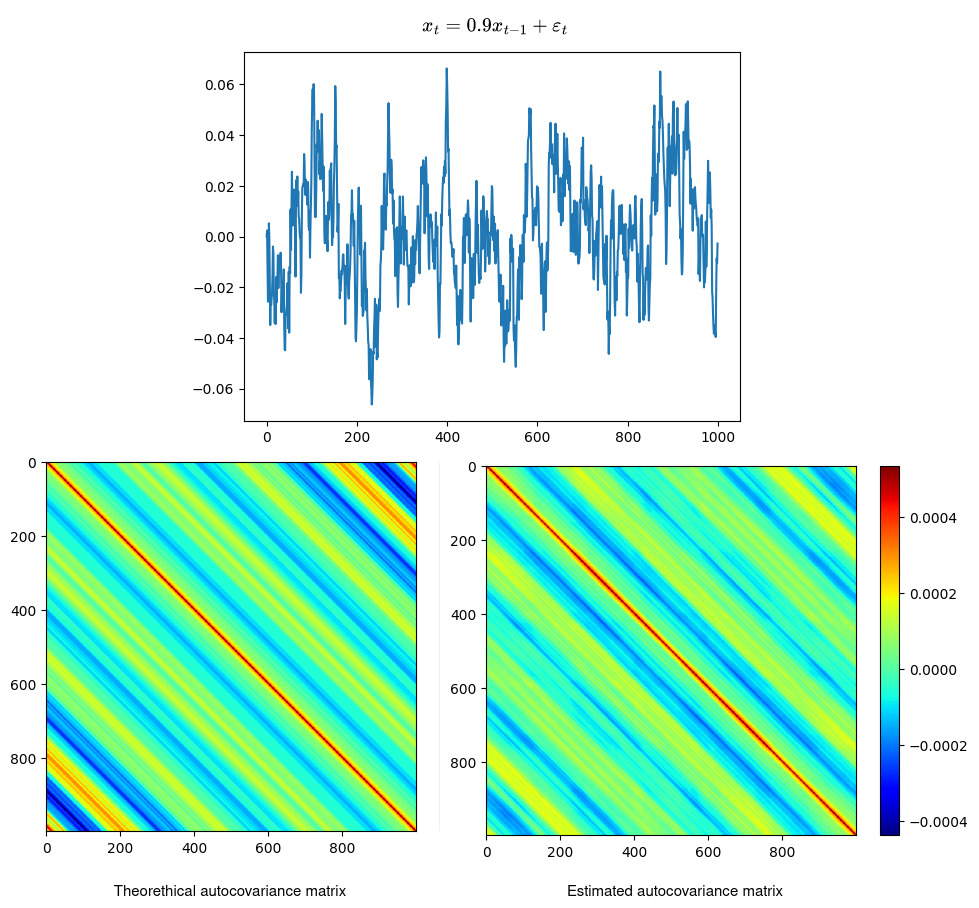}
		  \caption{Autocovariance matrix and estimated sliding autocovariance matrix.}\label{fig_windowed}
	\end{figure}

\end{example}
	
\subsection{Generalized least squares Granger causality test}

Going back to Granger causality test, in this section we present a novel Granger causality test based on wald tests and the estimation of the covariance 
matrix of the residuals via the sliding autocovariance matrix.

\begin{method}[\textbf{GLS Granger Causality test}]\label{meth_gls_granger}
Given two time series $\x$, $\y$, in order to assess whether $\x$ \textbf{causes} $\y$ \textbf{with lag} $L$,  we follow the following procedure, which is a variation 
of the classical Granger causality test:

\begin{enumerate}
\item Use OLS to obtain an estimate $(\boldsymbol \beta_{OLS}, \boldsymbol \beta_{OLS}')$
for the model 
\begin{equation}\label{eq_mod_gls_granger}
	y_t = \beta_0 + \sum^{p}_{k=1} \beta_k y_{t-k} + \sum^{p}_{k=1} \beta'_k x_{t-k} + \varepsilon_t
\end{equation}

\item Using the residuals of the previous model
\[r_t = y_t - \beta_0 - \sum^{p}_{k=1} \beta_k y_{t-k} - \sum^{p}_{k=1} \beta'_k x_{t-k} 
\]

estimate their covariance matrix using the sliding autocovariance matrix $\Omega_\tau$ as in (\ref{eq_sliding_autocov_matrix})

\item Use GLS and the previous covariance matrix to estimate again 
the parameters $(\boldsymbol \beta_{GLS}, \boldsymbol \beta_{GLS}')$ in the model (\ref{eq_mod_gls_granger}) as in (\ref{eq_gls}).

\item Use a Wald test with null hypothesis
\[
	H_0: \beta_{GLS\; k}' = 0 \text{ for all } k\leq p
\]
If the null hypothesis is rejected conclude that $\x$ \textbf{causes} $\y$ otherwise 
conclude the opposite.
\end{enumerate}
\end{method}

\section{Results}

We proceed now to assess the efficacy of the proposed GLS Granger causality test (\ref{meth_gls_granger}) in comparison with 
the classical Granger F-Test (sec. \ref{sec_granger_ftest}). 

\subsection{Simulated dataset}\label{sim_dataset}
Given a time series $\x$ we applied the following procedure to 
consistently generate a  
 \textbf{caused} series $\y$. The procedure consists of defining $\y$ in the following way

\begin{equation}\label{eq_sumulated_ts}
y_t = \sum^{L}_{k=1} x_{t-k} \cdot \beta_k + \varepsilon_t
\end{equation}

where $\beta_k$ are generated randomly and $\varepsilon$ is a time series that 
can be constructed in several ways depending in the type of causality 
that we want to simulate. We can consider:

\begin{enumerate}
\item[(M1)] $\varepsilon$ stationary time series, for instance a white noise   $\varepsilon_t \sim N(0,\sigma^2)$.

\item[(M2)] $\varepsilon$ is stationary with structural breaks, 
for instance considering $\varepsilon_t = \sim N(\mu_t, \sigma^2)$ with 
$\mu_t = 0$ for $t\leq t_b$ and $\mu_t = \mu$ for $t>t_b$.

\item[(M3)] $\varepsilon$ non-stationary time series with changing variance
for instance $\varepsilon_t \sim N(0,(t\cdot \sigma)^2)$.
\end{enumerate}

On the other hand, to simulate \textbf{non causality}, we will simply 
simulate two time series $\x$ and $\y$ by using auto-regressive processes 
with different parameters

\begin{align}
	x_t &= x_{t-1} \phi + \varepsilon_t\label{sim_non_causal}\tag{AR1}\\
	y_t &= y_{t-1} \phi' + \varepsilon_t\nonumber
\end{align}

\begin{observation} \label{obs_methods}
Notice that a regression model applied to predict $y$ from the 
lagged time series $B^k \y$ and  $B^k \y$ (using the backward operator as in obs. \ref{obs_backward}).
The regression parameters will approximate $\beta_1,\ldots \beta_L$ 
residuals of the regression will approximate $\varepsilon$.

With this in mind, (M1) will produce time series in which classical Granger F-test will 
be very effective. In contrast (M2) will give us stationary with structural breaks residuals and (M3) will
produce heteroskedastic residuals, therefore the classical Granger F-test will be less effective with these 
time series.

For this reason the introduced dataset (simulated using M1, M2 and M3) will be suitable 
for comparing the proposed GLS Granger causality test with 
the classical Granger F-Test.
\end{observation}

\begin{example}
	In the following figure \ref{fig_sim_data} we show 
	three examples of the generated dataset using 
	(M1), (M2) and (M3).
	
	Notice that in the graphs of figure \ref{fig_sim_data} we can observe 
	the causality, in the sense that changes $\x$ (shown in blue)
	cause changes in $\y$ after a number of lags.
	
	In the second graph the figure we see the structural
	break introduced in the residual of $\y$ (M2). 
	
	Finally, in the last graph of the figure  we appreciate the 
	residual with growing variance introduced by (M3) in $\y$. 

	\begin{figure}[H]
		\centering
		\includegraphics[scale=0.28]{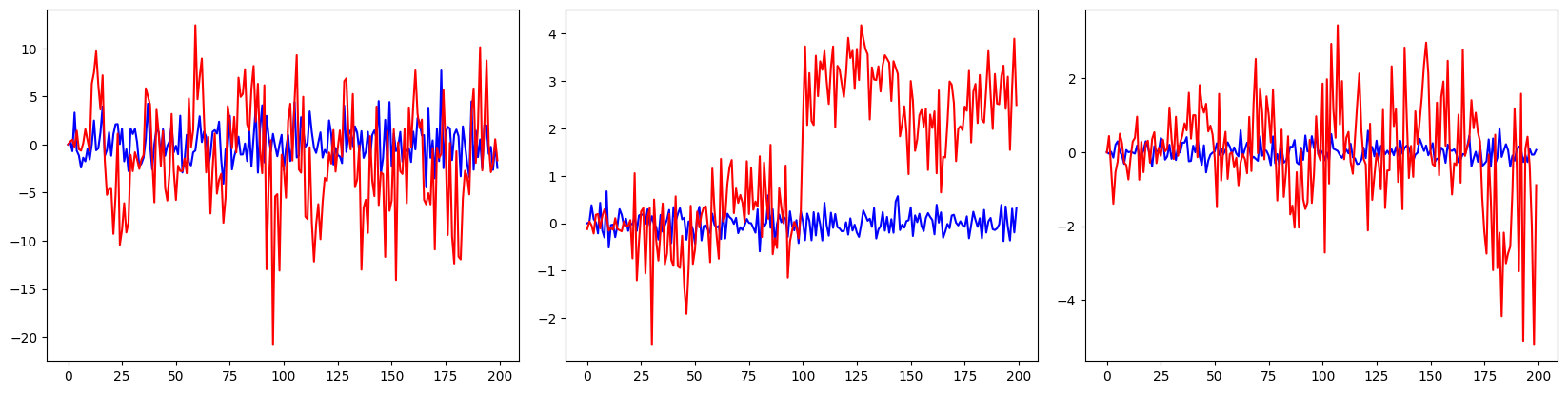}
		  \caption{Three pairs of time series generated using the previous method (for $L=15$ and random $\beta_1,\ldots,\beta_L$). In blue, the original time series 
		  $\x$ generated using an $AR(1)$ process. In red the \emph{caused} time series $\y$ 
		  generated using the three methods (M1), (M2) and (M3) respectively. }\label{fig_sim_data}
	\end{figure}
\end{example}

\subsubsection*{Experiment results}\label{sim_res_dataset}
We perform four experiments, 
each with 150 pairs of time series each with 600 points.
The lag used to simulate the causality is $L=15$.

The first 
three experiments consist on testing the performance of Classical Granger 
against the proposed GLS Granger by simulating causal relationships using methods
(M1), (M2) and (M3). 
For the sake of this comparison, we record the percentage of correct predictions by each method.

The last experiment attempts to search for \emph{false positives}. In this 
experiment we generate non-caused time series using the method (AR1) described before.

\begin{table}[H]
\centering
\begin{tabular}{lccc}
	\parbox{20ex}{Simulated causal\\ relationships} & \parbox{20ex}{simulation \\procedure} & \parbox{30ex}{$\%$ of correct classical\\ Granger F-test} & \parbox{20ex}{$\%$ of correct\\ GLS-Granger}\\
	\hline
	$y$ \textbf{caused by} $\x$ &(M1) stationary residual         &75.0\% &\textbf{96.6\%}\\
	$y$ \textbf{caused by} $\x$ &(M2) structural breaks residual  &57.3\% &\textbf{85.5\%}\\
	$y$ \textbf{caused by} $\x$ &(M3) heteroskedastic residual    &32.6\% &\textbf{42.6\%}\\
	$y$ \textbf{not caused by} $\x$ &(AR1)&\textbf{94.0\%}&\textbf{94.7}\%
\end{tabular}
\caption{Experiment results table}\label{tab_res2}
\end{table}

The window length $\tau$ used for the sliding autocovariance matrix estimation was $\tau = N/5$ where $N$ is the number of observations. 
This value was obtained using cross-validation, but greater values of $\tau$ produced very similar results.

In view of table \ref{tab_res2}, the proposed method gets more accurate results than the Granger F-test in every one of the datasets
simulated.

\subsubsection{Real  dataset}
Cryptocoins are known for their volatility and interdependence. Granger causality is a known tool to study the interdependence of 
cryptocoins, 
for instance \cite{kim2021causal} found a strong relationship between Bitcoin (BTN) and Ethereum (ETH) and \cite{yavuz2022causality} 
points out complex interdependence between the main cryptocoins.

We will use a dataset composed by the values of the main 10 cryptocurrencies 
(Bitcoin,Ethereum, Aave, BinanceCoin,  Cardano,   ChainLink, Cosmos, CryptocomCoin, Dogecoin, EOS, Iota,   Litecoin, Monero) from July 2020 to 
July 2021.

\begin{figure}[H]
	\centering
	\includegraphics[scale=0.32]{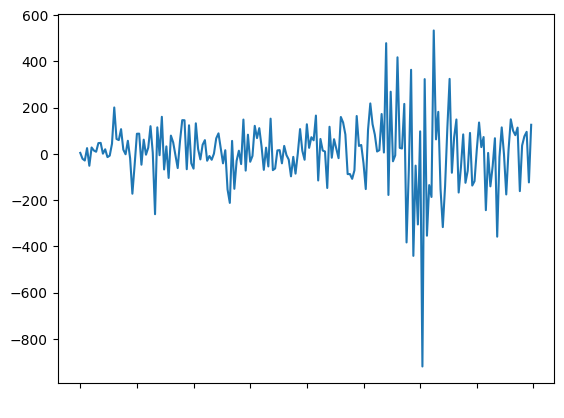}
	  \caption{Differentated data of the cryptocoin Ethereum from July 2020 to 
	  July 2021}\label{fig_eth_2022}
\end{figure}

The trend component of these time series was removed applying first order differentiation. Even after differentiation
a progressive change in variance is observed in the time series (see figure \ref{fig_eth_2022}). This suggests that even  though in some 
cases the series pass a Augmented Dickey-Fuller stationarity test, the OLS estimation performed in every Granger F-test will be imprecise 
or problematic. This situation has many similarities to the simulation preformed in (M3), for this reason our proposed method is more 
suitable to deal with the heteroskedastic behavior of the residuals.
(see sec. \ref{sim_dataset} and fig. \ref{fig_sim_data}).

We applied a Granger F-test and our proposed GLS Granger test on each pair of cryptocoins considered composing \emph{causals graphs}, i.e. 
a graph that has as nodes all the time series and as edges the causal relationships (if $\x$ causes $\y$ we draw $\x \to \y$)

We used the lag $L=1$, the optimal lag was obtained using Akaike Information Criterion (AIC).

\begin{figure}[H]
	\centering
	\includegraphics[scale=0.32]{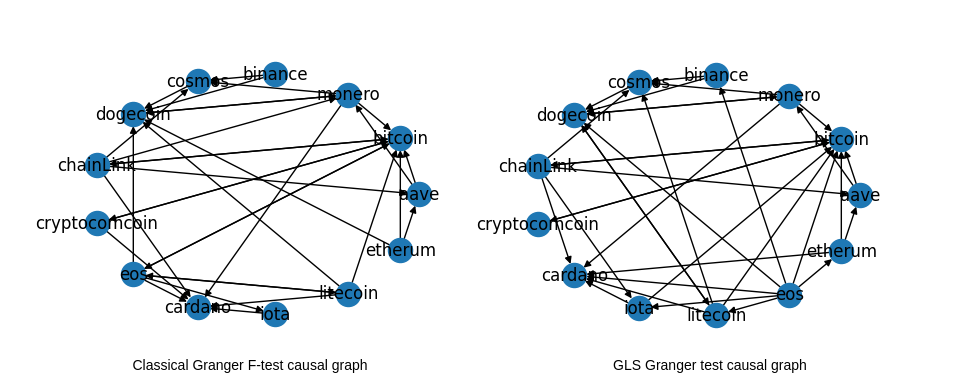}
	  \caption{Causal graphs}\label{fig_causal_crypto}
\end{figure}

The result is shown in Figure \ref{fig_causal_crypto}. The left graph of \ref{fig_causal_crypto} shows the Granger F-test causal graph, whereas the 
right graph show the resulting GLS Granger Graph. We obtained a very connected causal network as it is to expect from the behavior of cryptocurrencies.
Interestingly, the proposed method was able to capture more causal relationships, showing an even more connected network. 
It also noteworthy that the GLS Granger graph shows many causal relationship 
that connect the two leading cryptocoins Bitcoin and Ethereum with the rest of them. 
For instance the proposed method finds the relations Ethereum $\to$ cardano, 
Bitcoin $\to$ EOS, Bitcoin $\to$ ChainLink, Iota $\to$ Bitcoin. 

\section{Conclusions and Future work}

In this paper, we propose a generalization
of the Granger F-test to
uncover the temporal causal structures from heteroskedastic and structural-breaks time series trough the estimation of 
the residual autocovariance matrix and GLS. 

We demonstrate
its effectiveness on four simulation datasets and one real
application dataset. 

For future work, we are interested in researching other uses of the sliding covariance matrix in the field of time series classification
and machine learning.

\subsection*{Code availability}
Datasets and scripts for this article are available at github: \href{https://github.com/Granger-Causality-GLS}{https://github.com/Granger-Causality-GLS}

\bibliographystyle{amsalpha}

\bibliography{bibliography}

\providecommand{\bysame}{\leavevmode\hbox to3em{\hrulefill}\thinspace}
\providecommand{\MR}{\relax\ifhmode\unskip\space\fi MR }
\providecommand{\MRhref}[2]{%
  \href{http://www.ams.org/mathscinet-getitem?mr=#1}{#2}
}
\providecommand{\href}[2]{#2}
\begin{thebibliography}{KCP21}

\bibitem[Gra69]{granger1969investigating}
Clive~WJ Granger, \emph{Investigating causal relations by econometric models
  and cross-spectral methods}, Econometrica: journal of the Econometric Society
  (1969), 424--438.

\bibitem[Gra80]{granger1980testing}
\bysame, \emph{Testing for causality: a personal viewpoint}, Journal of
  Economic Dynamics and control \textbf{2} (1980), 329--352.

\bibitem[Gre03]{greene2003econometric}
William~H Greene, \emph{Econometric analysis}, Pearson Education India, 2003.

\bibitem[KCP21]{kim2021causal}
Myeong~Jun Kim, Nguyen~Phuc Canh, and Sung~Y Park, \emph{Causal relationship
  among cryptocurrencies: A conditional quantile approach}, Finance Research
  Letters \textbf{42} (2021), 101879.

\bibitem[LS01]{lee2001testing}
Junsoo Lee and Mark Strazicich, \emph{Testing the null of stationarity in the
  presence of a structural break}, Applied Economics Letters \textbf{8} (2001),
  no.~6, 377--382.

\bibitem[Per89]{perron1989great}
Pierre Perron, \emph{The great crash, the oil price shock, and the unit root
  hypothesis}, Econometrica: journal of the Econometric Society (1989),
  1361--1401.

\bibitem[SSS00]{shumway2000time}
Robert~H Shumway, David~S Stoffer, and David~S Stoffer, \emph{Time series
  analysis and its applications}, vol.~3, Springer, 2000.

\bibitem[Yav22]{yavuz2022causality}
G{\"U}L Yavuz, \emph{Causality and cointegration in cryptocurrency markets},
  Uluslararas{\i} {\.I}ktisadi ve {\.I}dari {\.I}ncelemeler Dergisi (2022),
  no.~34, 129--142.

\end{thebibliography}

\end{document}